# THE REDUCED MAXWELL-DUFFING DESCRIPTION OF EXTREMELY SHORT PULSES IN NON-RESONANT MEDIA

E.V. Kazantseva[a], A.I. Maimistov[a], J.-G. Caputo[b]

[a]Department of Solid State Physics, Moscow Engineering Physics Institute, Kashirskoye sh. 31, Moscow, 115409, Russia
[b]Laboratoire de Mathematiques, INSA de Rouen, B. P. 8, 76131 Mont-Saint-Aignan, Rouen, France
[b]Laboratoire de Physique theorique et modelisation, Universite de Cergy-Pontoise and C.N.R.S.

ABSTRACT

The propagation of extremely short pulses of electromagnetic field (electromagnetic spikes) is considered in the framework of a model where the material medium is represented by anharmonic oscillators with cubic nonlinearities (Duffing model) and waves can propagate only in the right direction. The system of reduced Maxwell-Duffing equations admits two families of exact analytical solutions in the form of solitary waves. These are bright spikes propagating on a zero background, and bright and dark spikes, propagating on a nonzero background. Direct simulations demonstrate that these pulses are very robust against perturbations. We find that a high frequency modulated electromagnetic pulse evolves into a breather-like one. Conversely a low frequency pulse transforms into a quasi-harmonic wave.

PACS: 42.65.Tg

Keywords: extremely short pulses, anharmonic oscillators, Duffing model, steady state pulse, soliton.

## 1. INTRODUCTION

At present extremely short pulses (ESP) of the electromagnetic field, which contain a few optical cycles, down to even half a cycle, attract a great deal of attention. Much work has already been done along this line of research [1-8]. As is customary, the description of the ESP evolution employed the total Maxwell equations without assuming a separation into the carrier wave and envelope. Generally, the Maxwell equations admit the propagation of electromagnetic waves in both directions. If however, the nonlinear contribution to the polarization of the medium is small, *the unidirectional* wave propagation may be assumed [4-10]. This approximation reduces the wave equation to a first-order one without any assumption about the shape of the waves. The unidirectional wave propagation approximation is frequently used for the simulation of ESP propagation in a homogeneous low density medium [11,12].

The nonlinear dynamics of the medium driven by the electromagnetic field can be modeled using anharmonic oscillators. In particular, the propagation of a linearly polarized ESP was considered [6,13] in the framework of the Duffing oscillators, the nonlinear response of the medium being cubic. This is the simplest generalization of the Lorentz model, which has been very useful to describe the propagation of an electromagnetic wave in a linear medium. Recently the Lorentz oscillator model was employed [12] to account for a linear retarded response of the medium and a nonlinear oscillator was considered to describe an instantaneous Kerr nonlinearity. The Duffing model takes into account the dispersion properties of both the linear and nonlinear responses of the medium so that it may represent better the nonlinear response on an electromagnetic pulse containing a few cycles.

The objective of the present work is to study the unidirectional propagation and interactions of linearly polarized ESPs in a nonlinear dispersive medium modeled by an anharmonic oscillator characterized by cubic nonlinearities. The paper is structured as follows. The model is derived in section 2. Dynamical invariants or integrals of motion are given in section 3. Two families of moving ESP solutions are found analytically in section 4 and confirmed using Hirota bilinear forms in section 5. The stability of the pulses and their collisions (for both signs of the polarity of the colliding pulses) are investigated numerically in section 6. We conclude in section 7.

## 2. THE REDUCED MAXWELL-DUFFING MODEL

The one-dimensional propagation of electromagnetic waves in a nonlinear medium is governed by the wave equation

$$\frac{\partial^2 E}{\partial z^2} + \frac{1}{c^2}\frac{\partial^2 E}{\partial t^2} = \frac{4\pi}{c^2}\frac{\partial^2 P}{\partial t^2}, \tag{1}$$

where $P$ is the polarization of the medium. According to the unidirectional wave approximation Eq. (1) can be replaced by the first-order equation

$$\frac{\partial E}{\partial z} + \frac{1}{c}\frac{\partial E}{\partial t} = -\frac{2\pi}{c}\frac{\partial P}{\partial t}. \tag{2}$$

We adopt a simple anharmonic-oscillator model for the medium, which is commonly used to approximate the medium response for an electromagnetic influence [14] (see also [15]). Here we will consider the oscillator with cubic anharmonicity. Here we will consider the oscillator with cubic anharmonicity. In addition we will assume the case of a homogeneous broadening medium, where all atoms have the same parameters. If $X$ represents the displacement of an electron from its equilibrium position, the equation of motion (which neglects friction) can be written as

$$\frac{\partial^2 X}{\partial t^2} + \omega_0^2 X + \kappa_3 X^3 = \frac{e_0}{m_{ef}} E(z,t) \tag{3}$$

where $\omega_0$ is an eigenfrequency of the oscillator, $\kappa_3$ is anharmonicity coefficients, $m_{ef} = 3m/(\varepsilon + 2)$ is effective mass of electron. Hereafter, we will use $m$ as a symbol for this effective mass. Finally, the dynamical variable $X$ is related to the medium polarization, $P = n_A e_0 X$, where $n_A$ is the density of the oscillators (atoms).

It is suitable the independent variables redenote as $t = z/l$, $x = \omega_0(t - z/c)$, and define the normalised dependent variables as

$$e = E/A_0, \text{ and } q = X/X_0, \tag{4}$$

where

$$A_0 = m_{ef}\omega_0^2 X_0 / e_0 = m_{ef}\omega_0^3 e_0^{-1}(2\mu/|\kappa_3|)^{1/2}, \quad X_0 = (2\mu\omega_0^2/|\kappa_3|)^{1/2},$$

$$l^{-1} = 2\pi n_A e_0^2 /(m_{ef}c\omega_0) = \omega_p^2 / 2c\omega_0, \tag{5}$$

and $\omega_p = (4\pi n_A e_0^2 / m_{ef})^{1/2}$ is the plasma frequency. In terms of the rescaled variables, equation (2) and (3) take the form

$$\frac{\partial e}{\partial t} = -\frac{\partial q}{\partial x}, \quad \frac{\partial^2 q}{\partial x^2} + q + 2\mu q^3 = e, \tag{6}$$

with the single remaining parameter $2\mu = k_3 X_0^2 / \omega_0^2$. The system of equations (6) furnish a final form of the model. In following we will referred them as *reduced Maxwell-Duffing equations* (RMD).

## 3. LAGRANGIAN AND INTEGRALS OF MOTION

The system of RMD equations can be derived as the Euler-Lagrange equations from the action functional

$$S = \int \mathsf{L}\,[q,\phi]dxdt$$

where the Lagrangian density is

$$\mathsf{L} = \frac{1}{2}\frac{\partial \phi}{\partial x}\frac{\partial \phi}{\partial t} + \frac{1}{2}\left(\frac{\partial q}{\partial x}\right)^2 - \frac{1}{2}q^2 - \frac{\mu}{2}q^4 + q\frac{\partial \phi}{\partial x}. \tag{7}$$

Application of the variational procedure to the action $S$ yields equations

$$\frac{\partial^2 \phi}{\partial t \partial x} = -\frac{\partial q}{\partial x}, \quad \frac{\partial^2 q}{\partial x^2} + q + 2\mu q^3 = \frac{\partial \phi}{\partial x}, \tag{8}$$

Identifying $\phi$ as a potential for the fields $q$ and $e$, so that $q = -\partial\phi/\partial t$ and $e = \partial\phi/\partial x$, makes these equations identical to the system of equations (6), which can be further transformed into the single equation

$$\frac{\partial q}{\partial t} + \frac{\partial q}{\partial x} + 6\mu q^2 \frac{\partial q}{\partial t} + \frac{\partial^3 q}{\partial x^2 \partial t} = 0. \tag{9}$$

From Lagrangian density (7) we can get the density of the canonical moments of the fields $\phi$ and $q$:

$$\pi_\phi(t,x) = \frac{\partial \mathsf{L}}{\partial \phi_{,t}} = (1/2)\phi_{,x}(t,x) = e(t,x), \quad \pi_q(t,x) = \frac{\partial \mathsf{L}}{\partial q_{,t}} = 0. \tag{10}$$

The density of the canonical Hamiltonian for this dynamical system can be obtained from L by means of the standard Legendre transformation,

$$\mathsf{H} = \frac{\partial \mathsf{L}}{\partial \phi_{,t}}\phi_{,t} + \frac{\partial \mathsf{L}}{\partial q_{,t}}q_{,t} - \mathsf{L} = -\frac{1}{2}\left(\frac{\partial q}{\partial x}\right)^2 + \frac{1}{2}q^2 + \frac{\mu}{2}q^4 - eq.$$

The variable e can be eliminated from it, using RMD equations, so that

$$\mathsf{H} = -\frac{\partial}{\partial x}\left(q\frac{\partial q}{\partial x}\right) + \frac{1}{2}\left(\frac{\partial q}{\partial x}\right)^2 - \frac{1}{2}q^2 - \frac{3\mu}{2}q^4. \tag{11}$$

Omitting the full derivative, the Hamiltonian corresponding to the density (11) takes the form

$$H = \int_{-\infty}^{+\infty}\left[\frac{1}{2}\left(\frac{\partial q}{\partial x}\right)^2 - \frac{1}{2}q^2 - \frac{3\mu}{2}q^4\right]dx.$$

The Hamiltonian is the first integral of motion of the RMB-equations. An additional integral of motion is the total canonical moment associated with the field $\phi$ that one may check on the basis of the RMD-equations (6).

$$I_1 = \int_{-\infty}^{+\infty}e(t,x)dx = \int_{-\infty}^{+\infty}\phi_{,x}(t,x)dx = \phi(t,x=\infty) - \phi(t,x=-\infty). \tag{12}$$

The magnitude of this integral is defined by boundary conditions only, thus it can be interpreted as topological charge in reduced Maxwell-Duffing model.

The third integral can be find by following. Using the canonical moment $\pi_\phi(x,t)$ one can rewrite the equations (6) as

$$\frac{\partial \pi_\phi}{\partial t} = -\frac{\partial q}{\partial x}, \quad \pi_\phi = \frac{1}{2}\left(\frac{\partial^2 q}{\partial x^2} + q + 2\mu q^3\right).$$

From the first equation of this system it follows $\pi_\phi \pi_{\phi,t} = -(1/2)\pi_\phi q_{,x}$. Taking into account second equation one can obtain the expression

$$(\pi_\phi^2)_{,t} = -(1/4)\left(q_{,x}^2 + q^2 + 2\mu q^4\right)_{,x}.$$

Thus one get third integral of motion

$$I_3 = -\int_{-\infty}^{\infty}\pi_\phi^2(x,t)dx = \frac{1}{4}\int_{-\infty}^{\infty}\left\{q + 2\mu q^3 + q_{,xx}\right\}^2 dx. \tag{13}$$

Taking into account the relation (10), this integral may be interpreted as a "pulse energy",

$$I_3 = (1/4) \int_{-\infty}^{\infty} e^2(t,x) dx. \tag{14}$$

It should be point that the Lagrangian of RMD model is an example of a degenerated Lagrangian system. The expressions in (10) indicate that this Lagrangian leads to constrained Hamiltonian system, where $\pi_\phi(t,x) = (1/2)\phi_{,x}(t,x)$ and $\pi_q(\zeta,\tau) = 0$ represent the primary constraint [16]. The conservation of the total canonical moment (12) corresponds to the invariance of the system under consideration with respect to shift of the field $\phi$ for a constant. It is not space translation symmetry, as it usually occurs, when referring to the moment.

To consider the space-time translation symmetry of the RMD model it is suitable to denote new variables: $y_1 = t$, $y_2 = x$ and $u_1 = \phi, u_2 = q$. For any system, if the space-time variables are not explicitly included in the Lagrangian, there are the conservation laws of the form

$$\sum_k \frac{\partial}{\partial y_k} T_i^k = 0, \tag{15}$$

where the energy-moment tensor $T_i^k$ is denoted as

$$T_i^k = \sum_a \frac{\partial L}{\partial u_{a,k}} u_{a,i} - \delta_{ki} L.$$

In the case of the RMD model we have two integrals of motion resulting from (15)

$$Q_1 = \int_{-\infty}^{+\infty} T_1^1 dx, \quad Q_2 = \int_{-\infty}^{+\infty} T_2^1 dx. \tag{16}$$

Bu using the Lagrangian (7) one can find

$$T_1^1 = -\frac{1}{2}\left(\frac{\partial q}{\partial x}\right)^2 + \frac{1}{2}q^2 + \frac{\mu}{2}q^4 - q\frac{\partial \phi}{\partial x}, \quad T_2^1 = -\frac{1}{2}\left(\frac{\partial \phi}{\partial x}\right)^2.$$

The substitution of these expressions into integrals of (16) leads to $Q_1 = H$, $Q_2 = -(1/2)I_2$. Thus we obtain the interpretation integrals $I_2$ and $H$ as the total moment and total energy of in the RMD model. Unlike the total canonical moment $I_1$, the total moment $Q_2$ reflects invariance of the RMD model with respect to space translation.

## 4. ANALYTICAL SOLUTIONS FOR THE EXTREMELY SHORT PULSES

It seems plausible that the system of RMD equations is not integrable. Nevertheless, some exact analytical solutions, describing the propagation of ESPs without destruction, can be found. In order to get these steady state solutions, one should assume that e and $q$ depend on a single variable,

$$\eta = x - t/\alpha = \omega_0(t - z/V), \tag{17}$$

with some constant $\alpha$. An expression for the velocity $V$ of a steadily moving pulse then follows from Eq. (4),

$$V^{-1} = c^{-1}\left[1 + (1/2\alpha)(\omega_p/\omega_0)^2\right]. \tag{18}$$

Hence, parameter $\alpha$ defines the velocity of the steady state ESP and we should get one parametric family of exact analytical solutions of the RMD equations. In general, the chousing the boundary conditions results in different solutions of these equations. Here we restrict our attention to solitary wave solutions.

**4.1 Steady state pulse on zero background**

Let us consider the following initial and boundary conditions

$$e(t=0, x) = e_0(x), \tag{19}$$

and

$$e_0(x) = 0, \quad q(t,x) = \partial q(t,x)/\partial x = 0 \text{ at } x \to \pm\infty.$$

The first equation of the system (6) can be integrated to yield

$$e = \alpha q. \tag{20}$$

Next, the second equation from the system (6) takes the form

$$\frac{d^2 q}{d\eta^2} + (1-\alpha)q + 2\mu q^3 = 0. \tag{21}$$

If $\alpha > 1$ and $\mu > 0$, this equation has a family of exact solutions parameterized by the continuous *positive* parameter $\alpha - 1$ and discrete one,

$$q(x,t) = \pm\sqrt{(\alpha-1)/\mu}\,\text{sech}\left[\sqrt{(\alpha-1)}(x - t/\alpha - x_0)\right]. \tag{22}$$

$$e(x,t) = \pm\alpha\sqrt{(\alpha-1)/\mu}\,\text{sech}\left[\sqrt{(\alpha-1)}(x - t/\alpha - x_0)\right] \tag{23}$$

The expression (23) corresponds to the one spike of the electromagnetic field, propagating without form distortion in a non-resonant medium with cubic nonlinearity.

**4.2 Steady state pulse on non-zero background**

If the medium is preliminary polarized by continuum electric field, the oscillator coordinate is shifted from the equilibrium position (i.e., atoms have a constant electronic polarizability induced by an external electric field). Let us denote this new position as $q_0$. The initial and boundary conditions can be written as:

$$e(t=0, x) = e_0(x),\tag{24}$$

and

$$e_0 = q_0 + 2\mu q_0^3,\ q(t,x) = q_0 \neq 0,\ \partial q(t,x)/\partial x = 0\ \text{at}\ x \to \pm\infty.$$

We introduce the new variables $f = q - q_0$ and $u = e - e_0$, which approach zero at infinite. That results to the following equations

$$\frac{\partial u}{\partial t} = -\frac{\partial f}{\partial x},\quad \frac{\partial^2 f}{\partial x^2} + f + 2\mu(3q_0^2 f + 3q_0 f^2 + f^3) = u.\tag{25}$$

Looking for a steady state solution of (25), we obtain

$$\frac{d^2 f}{d\eta^2} + (1-\alpha_1)f + 6\mu q_0 f^2 + 2\mu f^3 = 0.\tag{26}$$

where $\alpha_1 = \alpha - 6\mu q_0^2$. Integrating this equation once, taking into account zero boundary conditions, one gets

$$(df/d\eta)^2 = (\alpha_1 - 1)f^2 - 4\mu q_0 f^3 - \mu f^4.\tag{27}$$

The substitution of $f = 1/y$ transforms this into the following equation

$$(dy/d\eta)^2 = (\alpha_1 - 1)\left[(y - y_0)^2 - \Delta^2\right],$$

where

$$y_0 = 2\mu q_0(\alpha_1 - 1)^{-1},\ \Delta^2 = \mu(\alpha_1 - 1)^{-1} + y_0^2.\tag{28}$$

The substitution $y = y_0 \pm \Delta \cosh\varphi$ reduces this equation to the trivial one $d\varphi/d\eta = (\alpha_1 - 1)^{1/2}$. Thus we have the solution of equation (27) written as

$$f(t,x) = \left[y_0 \pm \Delta \cosh\{\sqrt{(\alpha_1 - 1)}(\eta - \eta_0)\}\right]^{-1}\tag{29}$$

If $q_0 \to 0$ this expression reproduces the formula (22). The electric field of the electromagnetic wave is given by the following expression:

$$e(t,x) = e_0 + \alpha\left[y_0 \pm \Delta \cosh\{\sqrt{(\alpha_1 - 1)}(\eta - \eta_0)\}\right]^{-1} =$$
$$= e_0 + \frac{\alpha(\alpha_1 - 1)}{2\mu q_0 \pm \sqrt{\mu(\alpha_1 - 1) + 4\mu^2 q_0^2}\cosh\{\sqrt{(\alpha_1 - 1)}(\eta - \eta_0)\}}\tag{30}$$

The plus sign in (30) corresponds to a bright spike of the electromagnetic field on a constant background; the minus sign in (30) corresponds to a dark solitary wave: the narrow hole propagating on background. It is a new type of steady state solutions of the RMD equations.

Now we can consider the special case when parameter $\alpha_1$ is equal unit. Then equation (27) takes the form

$$(df/d\eta)^2 = -4\mu q_0 f^3 - \mu f^4 .. \tag{31}$$

Substitution $f = -1/y$ transforms it into $(dy/d\eta)^2 = \mu(4q_0 y - 1)$, the integral of which leads to the following expression for $f$

$$f(x,t) = \frac{-4q_0}{1 + 4\mu q_0^2 (x - t/\alpha - x_0)^2}. \tag{32}$$

Here $x_0$ is the constant of integration, which indicates the location of the maximum of the steady state dark pulse. One may name this pulse "algebraic soliton", due to its decay rate with time and coordinate. For this solution we have the following relations

$$q(x,t) = q_0 - \frac{4q_0}{1 + 4\mu q_0^2 (x - t/\alpha - x_0)^2}, \tag{33}$$

and

$$e(x,t) = q_0 + 2\mu q_0^3 - \frac{4q_0(1 + 6\mu q^2)}{1 + 4\mu q_0^2 (x - t/\alpha - x_0)^2}. \tag{34}$$

Note that when the initial medium state $q_0$ is large, the amplitude of the electric field is much larger than the medium variable. For $q_0 \gg 1$ we have a dark solitary wave having a bright spot in the center while for $q_0 \ll 1$ we have a solitary wave superimposed on a nonzero background..

## 5. BILINEAR FORM OF THE REDUCED MAXWELL-DUFFING EQUATIONS

If the following substitutions

$$e = \frac{a}{h}, \quad q = \frac{b}{h}, \tag{35}$$

are used, then the equations (6) can be rewritten as

$$\frac{1}{h^2} D_t(a \cdot h) + \frac{1}{h^2} D_x(b \cdot h) = 0, \qquad (36.1)$$

$$\frac{1}{h^2} D_x^2(b \cdot h) - \frac{b}{h^3} D_x^2(h \cdot h) + \frac{b}{h} + 2\mu \frac{b^3}{h^3} = \frac{a}{h}, \qquad (36.2)$$

where Hirota's D-operators $D_x(a \cdot b) \equiv (\partial_x a)b - a(\partial_x b)$ [17], and so on, were used. To derive the last equation we have followed the rule

$$\frac{\partial^2}{\partial x^2}\left(\frac{b}{h}\right) = \frac{1}{h^2} D_x^2(b \cdot h) - \frac{b}{h^3} D_x^2(h \cdot h). \qquad (37)$$

Multiplying the first equation by $h^2$, the second equation by $h^3$ and collecting the terms of equal order of $h^{-1}$, we can rewrite the resulting equations as a system of bilinear ones

$$D_t(a \cdot h) + D_x(b \cdot h) = 0, \qquad (38.1)$$
$$D_x^2(b \cdot h) = ah - bh, \qquad (38.2)$$
$$D_x^2(h \cdot h) = 2\mu b^2. \qquad (38.3)$$

Let us do the usual assuming to solve the equations (38) [17,18]:

$$a = \alpha \exp(\theta), \quad b = \beta \exp(\theta), \quad h = 1 + h_1 \exp(\theta) + h_2 \exp(2\theta),$$

where $\theta = kx - \Omega t$ [17,18]. The substitution of these expressions into (38) results in algebraic equations with respect of $\exp(\theta)$. Equating the coefficients of the different powers of $\exp(\theta)$ to zero, one can obtain the system of equations defining $a$, $b$, $h_1$ $h_2$. From (38.1) we get $\alpha\Omega - \beta k = 0$. From (38.2) the two conditions follow: $h_1 = 0$ and $\beta k^2 = \alpha - \beta$. These expressions lead to the "dispersion relation"

$$\Omega = k/(1 + k^2). \qquad (39)$$

Equation (38.3) yields three relations: $h_1 k^2 = 0$, $8h_2 k^2 = \beta^2$, $h_1 h_2 k^2 = 0$. As $h_1 = 0$, we have only second relation defining $h_2$, i.e., $h_2 = (1/8)(\beta/k)^2$. Thus, we find a one-soliton solution of the bilinear equations (38)

$$a = (1 + k^2)\beta \exp(\theta), \quad b = \beta \exp(\theta), \quad h = 1 + (\beta^2/8k^2)\exp(2\theta).$$

These relations yield the solution of the equations (6), which is correlated with the steady state one that was obtained earlier:

$$e = \frac{(1 + k^2)\beta \exp(\theta)}{1 + (\beta^2/8k^2)\exp(2\theta)}, \qquad q = \frac{\beta \exp(\theta)}{1 + (\beta^2/8k^2)\exp(2\theta)}. \qquad (40)$$

It would be interesting to find the two-soliton solution of the bilinear equations (38). However the standard approach [17, 18] is not successful. This may indicate that there are not two- and more- soliton solutions of the system of RMD equations.

## 6. NUMERICAL SIMULATION OF THE PROPAGATION AND COLLISIONS OF THE PULSES

Now that we have obtained analytical solutions corresponding to the solitary waves, it is important to test their stability and interactions using numerical simulations. For this purpose we write the system (6) in the following form:

$$\frac{\partial e}{\partial t} = -p, \qquad (41)$$

$$\frac{\partial q}{\partial x} = p, \quad \frac{\partial p}{\partial x} = e - q - 2\mu q^3. \qquad (42)$$

One can use any technique of numerical integration of ordinary differential equations to solve this system both in $x$ and $t$. We employed the fourth-order Runge-Kutta routine. As initial conditions, we took the analytical solutions given by Eqs. (23) and (30) at $t=0$. The boundary conditions are $q(t,x) = q_0, \partial q(t,x)/\partial x = 0$ at $x \to \pm\infty$.

The propagation and interaction of the steady state pulses in the dispersive medium with both quadratic and cubic nonlinearities were thoroughly studied in [3,4]. There, direct simulations demonstrated the strong stability of the pulses against various perturbations. Collisions between pulses show that these interact nearly elastically, irrespective of their relative polarity, unless their relative velocity is very small. In this last case collisions are inelastic, resulting in the generation of radiation and a new small-amplitude soliton.

Here we found two types of steady state pulses: an electromagnetic spike propagating in a ground state medium and a pulse propagating in a medium polarized by a constant electric field. In this last case the electromagnetic pulse propagates steadily over the constant electric background. Hence, the stability of the steady state pulses of the cubic nonlinear medium should be considered separately for both cases.

### 6.1 Propagation of pulses on a zero background

In this subsection we consider the stability of the steady state solutions corresponding to spikes propagating in non-polarized media, where the oscillators are initially in the ground state. In this case the initial and boundary conditions in the numerical simulations have been chosen as (19) with $e_0(x)$ defined by the expression (23). Below the parameter $\mu$ is set to 0.3 unless otherwise indicated.

As the velocities of the pulses are determined by their widths, the angle of the trajectories in the ($x,t$) plane and therefore the interaction time of the pulses can be modified by choosing their initial widths.

In the case of equal polarities if the relative velocities of the colliding pulses are considerably different, the pulses do not change their form and velocity after interaction. Figure 1 shows the collision of two bright spike solutions of (23) with $\alpha = 2$ and $\alpha = 4$. Strictly speaking the solitary wave solutions of the model under consideration are not solitons since the system of equations (6) has no Painlev'e properties [19], but numerical simulations prove that these steady state pulses collide almost elastically.

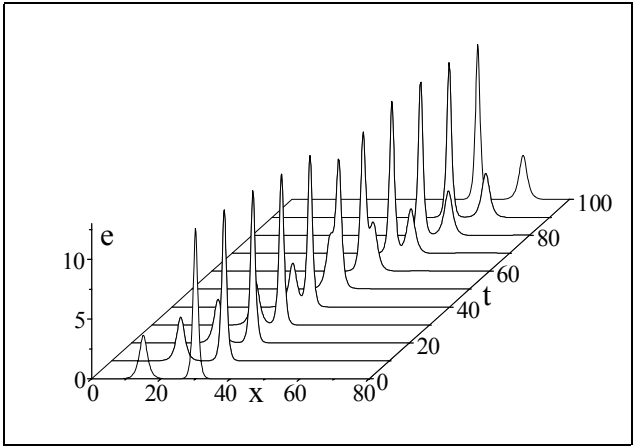

*Fig.* 1

Collision between two bright steady state pulses on a zero background. Their relative velocities differ sufficiently and the interaction does not change their shapes. The parameter $\mu = 1/3$

When decreasing the relative velocities of the colliding pulses, which are chosen with equal polarities, a strong mutual energy exchange takes place and the two colliding pulses never completely overlap. Fig. 2 is a typical picture illustrating this result, which is quite similar to the classical description of collisions between solitons in the modified Korteweg - de Vries equation [18].

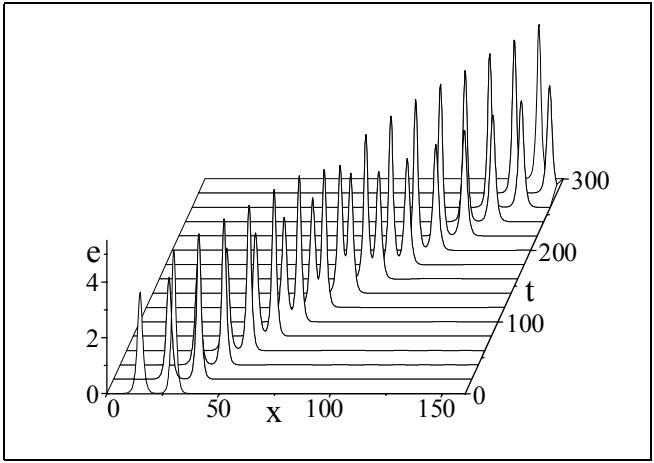

*Fig.* 2.

Energy exchange between two bright pulses traveling at very close velocities ($\alpha = 2$ and $\alpha = 2.4$).

In [4] the analysis of the collision between pulses in the quadratic-cubic nonlinear medium showed that the elasticity of interaction depends on the polarities of the pulses under consideration. Here this is not the case. As long as the amplitudes of the colliding pulses are different, the interaction is almost elastic irrespective of the polarity of the pulses, a fact demonstrated by fig. 3. When the relative velocity of the pulses is now reduced, again a strong mutual energy exchange takes place as shown in fig. 4.

To summarize this study of the collisions between two spikes, on can say that the result of the collision of the two spikes depends on the difference of the pulse widths. The two spikes can penetrate through each other without appreciable changes, if their widths are different enough. But the collision leads to more interesting results when the pulses velocities are close. In this case as seen from fig. 3 the energy due to the decay of the smaller pulse (irrespective of it's polarity) is transferred to the larger pulse and a small solitary wave appears during the interaction process.

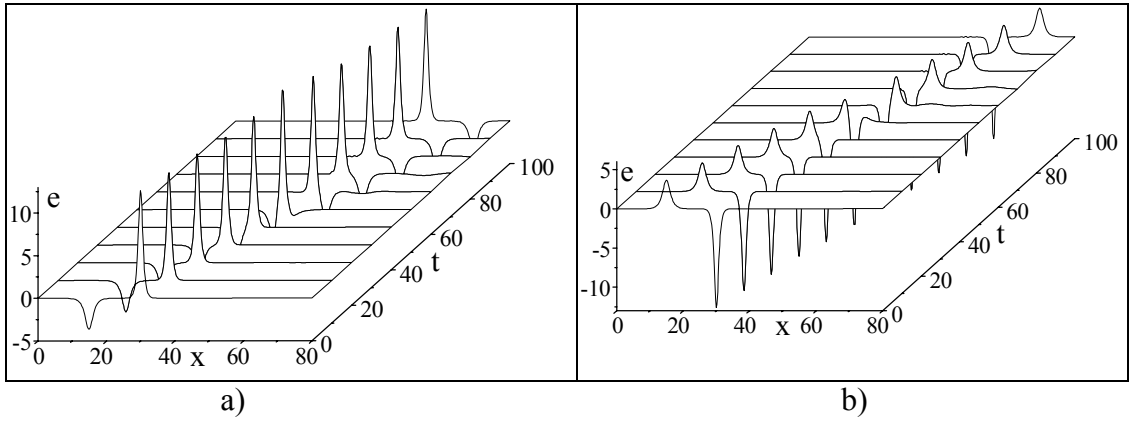

*Fig.* 3. Collision between pulses with negative and positive polarities. In fig. 3a the pulse of negative polarity is such that $\alpha = 2$ while the pulse with positive polarity is such that $\alpha = 4$. This is the converse in fig. 3b.

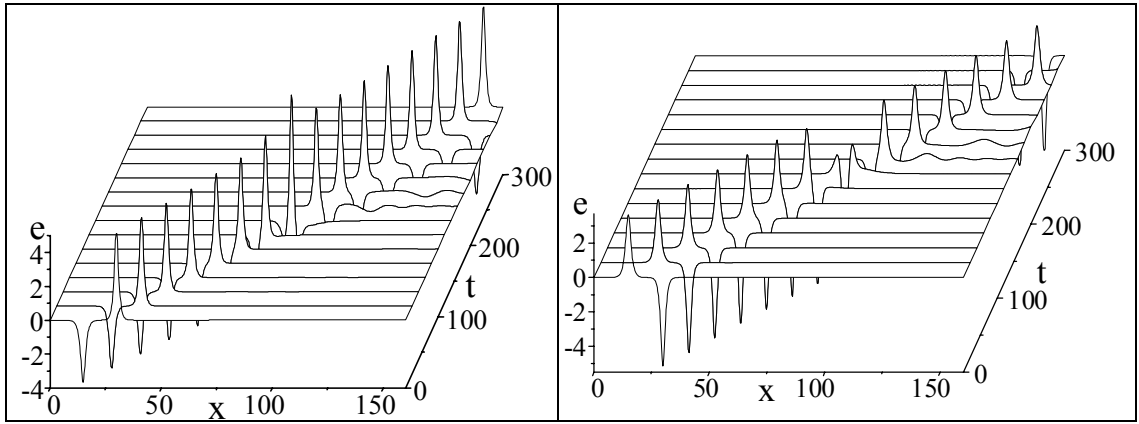

*Fig.* 4. Collision between pulses with negative and positive polarities that have very close velocities. In fig. 4a the pulse of negative polarity is such that $\alpha = 2$ while the pulse with positive polarity corresponds to $\alpha = 2.4$. This is the converse in fig. 4b.

The stability of the pulses under weak perturbations is also very interesting. We find that the steady state pulses under consideration appear to be stable in this sense. An example is presented in fig.5 which shows the evolution of a bright steady state pulse for $\alpha = 2$ initially perturbed by a harmonic wave packet, i.e.,

$$e_0(x) = 2\sqrt{1/0.3}\,\text{sech}(x-15) + 0.5\cos(5x)[\vartheta(x-30) - \vartheta(x-40)],$$

where $\vartheta(x)$ is the Heaviside step-like function: $\vartheta(x) = 0$ if $x<0$, and $\vartheta(x) = 1$ if $x>0$.

Increasing the initial pulse energy beyond the energy of the steady state pulse can result in its decay to one steady state pulse or more (depending on the initial pulse energy) and radiation. The formation of a steady state pulse from an arbitrary initial pulse also proves the stability of such solitary waves. Fig. 6. shows this phenomenon.

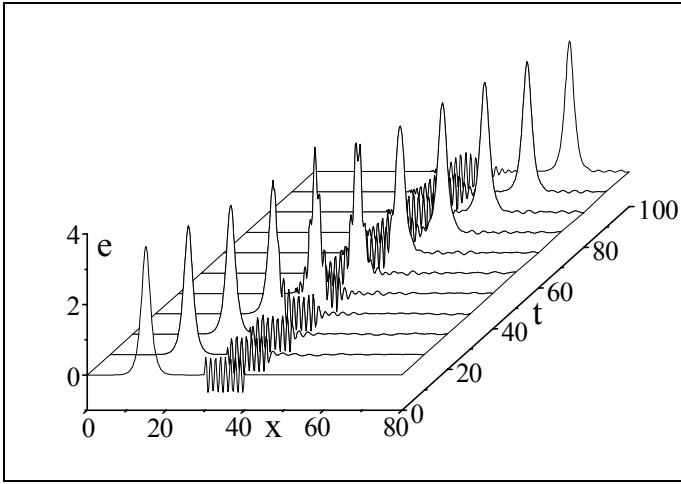

*Fig.5.*

Evolution of the pulse with $\alpha = 2$ initially perturbed by the addition of the harmonic wave packet $0.5\cos(5x)$.

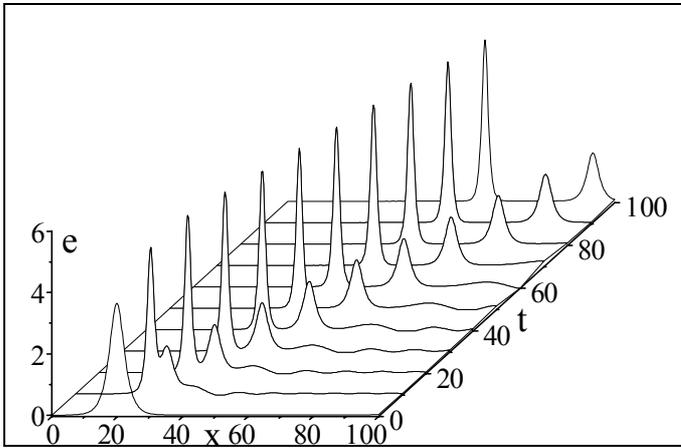

*Fig.6.*

Decay of a high-energy initial pulse $e(x, t=0) = (2/\sqrt{\mu})\operatorname{sech}(0.5x)$, which leads to a steady state pulse together with some radiation.

To conclude, our investigation demonstrates that the steady state solutions of the RMD equations on a zero background behave as solitons of completely integrable models.

**6.2 Pulse propagation on a non-zero background**

In the absence of an electromagnetic wave the state of the medium polarized by constant electric field is stable. It can be shown that modulational instability is absent in this model. Small perturbations of the background are not amplified. Now let us consider the stability of the steady state electromagnetic spikes we found propagating on such a background. The initial and boundary conditions have been chosen as (24), where the initial solitary pulse has the following form:

$$e(t=0, x) = e_0 + \alpha \left[ y_0 \pm \Delta \cosh\left\{ \sqrt{(\alpha_1 - 1)}(\eta - \eta_0) \right\} \right]^{-1}$$

with $\Delta$ defined by (28).

As in the case of zero background the perturbation of a steady state pulse by a weak harmonic wave does not result in the destruction of this spike, at the same time the harmonic wave packet transforms into a dispersive wave (fig. 7.). The background is destroyed neither by a weak harmonic wave (fig.7.) nor by a strong perturbation such as a high-energy pulse that breaks up into solitary waves (fig. 8.).

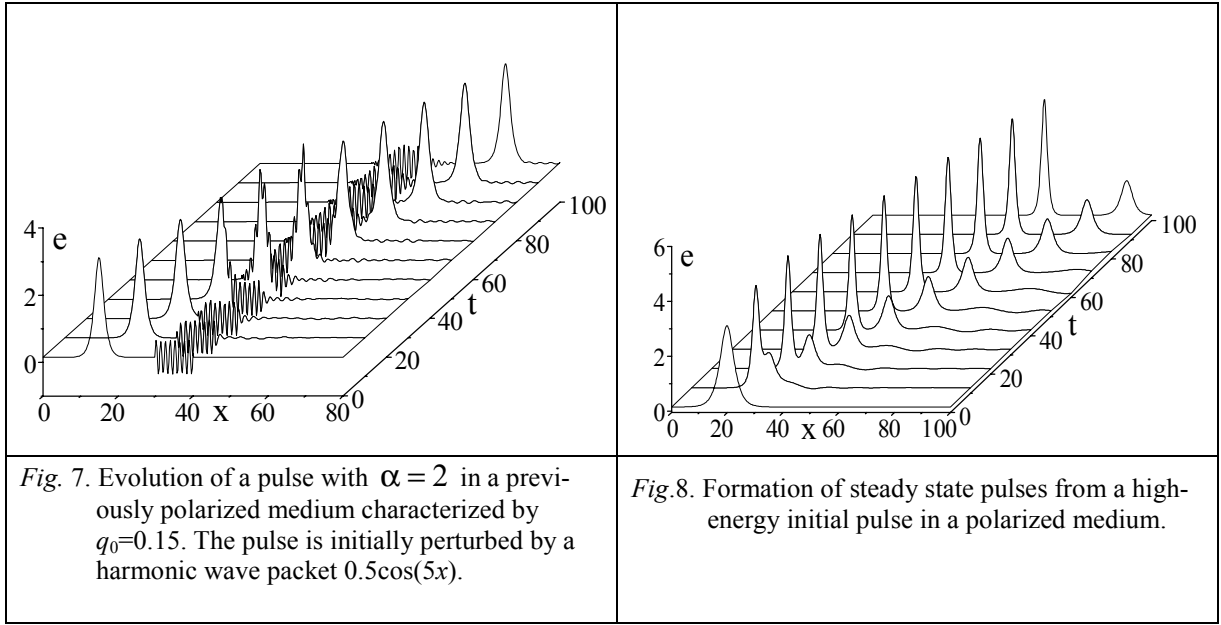

*Fig.* 7. Evolution of a pulse with $\alpha = 2$ in a previously polarized medium characterized by $q_0$=0.15. The pulse is initially perturbed by a harmonic wave packet 0.5cos(5$x$).

*Fig.*8. Formation of steady state pulses from a high-energy initial pulse in a polarized medium.

To investigate the stability of the solitary waves (30) over a background under strong perturbations, collisions of these steady state pulses were simulated. It was found that the pulses with sufficiently different amplitudes interact almost like solitons. However a weak radiation is emitted after collision which indicates the nonelasticity of the interaction. This result does not depend on the relative polarity of the colliding pulses.

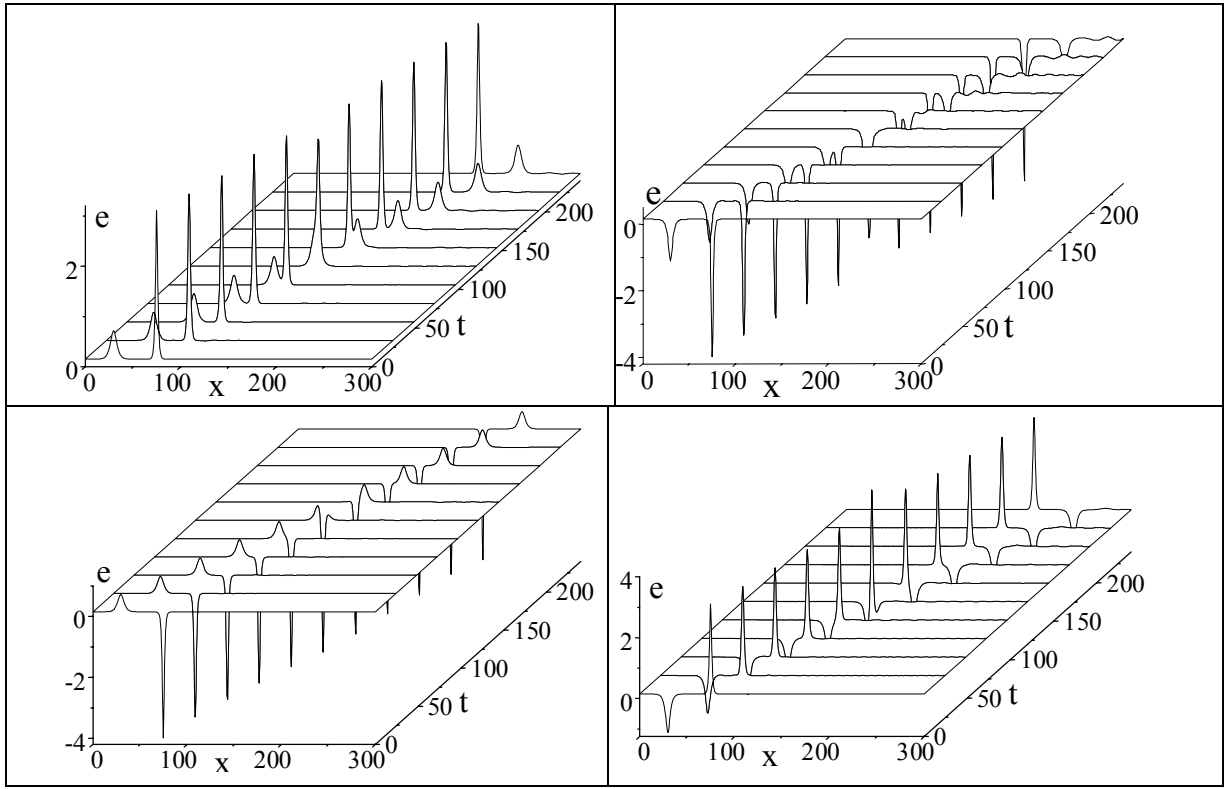

*Fig.* 9 Interactions between steady state pulses on a background corresponding to $q_0$=0.15 with different polarities. The faster pulse is such that $\alpha_1$=1.2 and the slower one is such that $\alpha_1 = 2$.

### 6.3 Breather-like pulses on a zero background

Equation (9) can be represented in the form of the modified Korteweg-de Vries (mKdV) equation with the additional term:

$$\frac{\partial q}{\partial t} + \frac{\partial q}{\partial x} - 6\mu q^2 \frac{\partial q}{\partial x} - \frac{\partial^3 q}{\partial x^3} = R[q], \qquad (43)$$

where

$$R[q] = -\frac{\partial^2}{\partial x^2}\left(\frac{\partial q}{\partial t} + \frac{\partial q}{\partial x}\right) - 6\mu q^2 \left(\frac{\partial q}{\partial t} + \frac{\partial q}{\partial x}\right)$$

is the difference between the RMD and mKdV equations. Thus a good agreement between the solutions of these two models can be expected if $q_{,x} \approx -q_{,t}$, hence the first equation of the system (6) gives the following: $e \approx q$, and (9) lead to

$$\frac{\partial e}{\partial t} + \frac{\partial e}{\partial x} - 6\mu e^2 \frac{\partial e}{\partial x} - \frac{\partial^3 e}{\partial x^3} = R[e].$$

The possible occurrence of breather-like pulses in the RMD model was an objective of our numerical investigation. Since the evolution equations of the RMD and mKdV models bear a great resemblance to each other and the numerical simulation of the steady state pulses of the RMD model shows significant stability both during collisions and perturbations, it seems natural to consider a breather solution of the mKdV equation as initial condition for the RMD equations

$$e(t=0,x) = -\frac{4\beta}{\alpha_1} \frac{\alpha_1 \cos\theta_2 \operatorname{ch}\theta_1 - \beta\operatorname{sh}\theta_1 \sin\theta_2}{\operatorname{ch}^2 \theta_1 + (\beta/\alpha_1)^2 \sin^2 \theta_2} \qquad (44)$$

with

$$\theta_1 = 2\beta(x-x_{10}) + 8\beta(\beta^2 - 3\alpha_1^2 - 0.25)t,$$
$$\theta_2 = 2\alpha_1(x-x_{20}) + 8\alpha_1(\alpha^2 - 3\beta^2 + 0.25)t$$

In the numerical simulation we use as initial condition for the RMD model a breather solution of the mKdV equation given by (44). We will consider the stability of these pulses depending on their frequency. In the following simulation $\beta$ is set to 0.5, and $\alpha_1$ is varied.

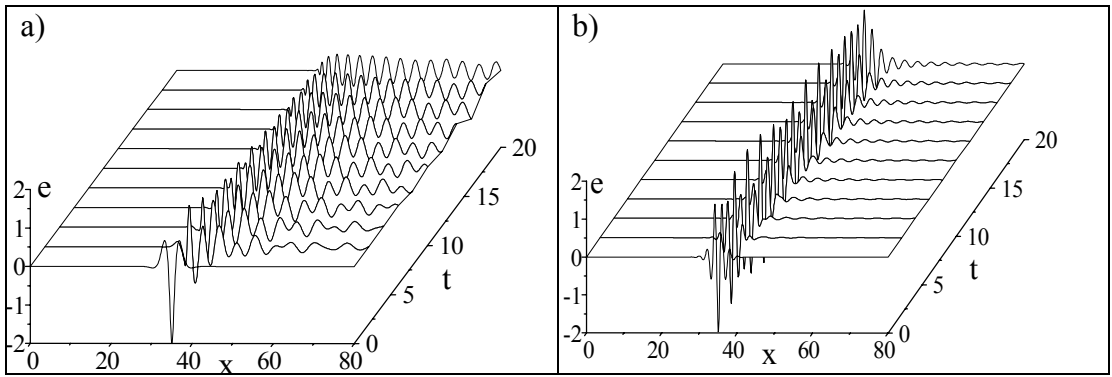

*Fig.* 10. A breather solution of the mKdV model in the RMD model: (a) corresponds to $\alpha_1 = 0.5$, while (b) corresponds to $\alpha_1 = 1.5$.

As seen from fig.10, a low-frequency pulse is broadened due to dispersion and decays into quasiperiodic waves. Localized breather-like pulses do not form in this case. Increasing the initial pulse frequency leads to pulse stabilization, as fig. 11 demonstrates.

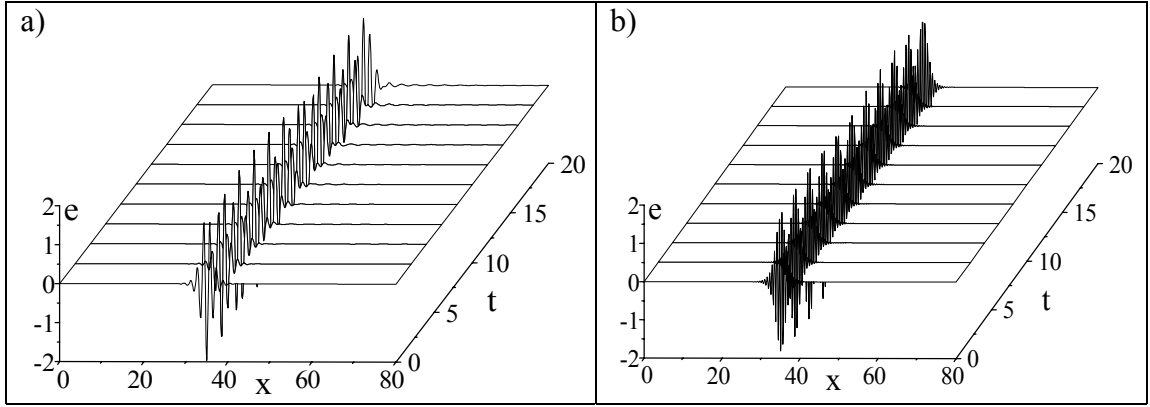

*Fig.* 11. A high-frequency pulse preserves its shape over long distances. (a) $\alpha_1=2$, and (b) - $\alpha_1=5$.

Collisions of such high-frequency pulses with a steady state pulse of the RMD model (fig. 12) also demonstrate the stability of these breather-like pulses, thought they are not breathers of the model under consideration. In fig. 12a $\alpha_1=3$, and in fig. 12b $\alpha_1=5$. The steady state pulse is characterized by $\alpha_1=3$ with $\mu=0.1$.

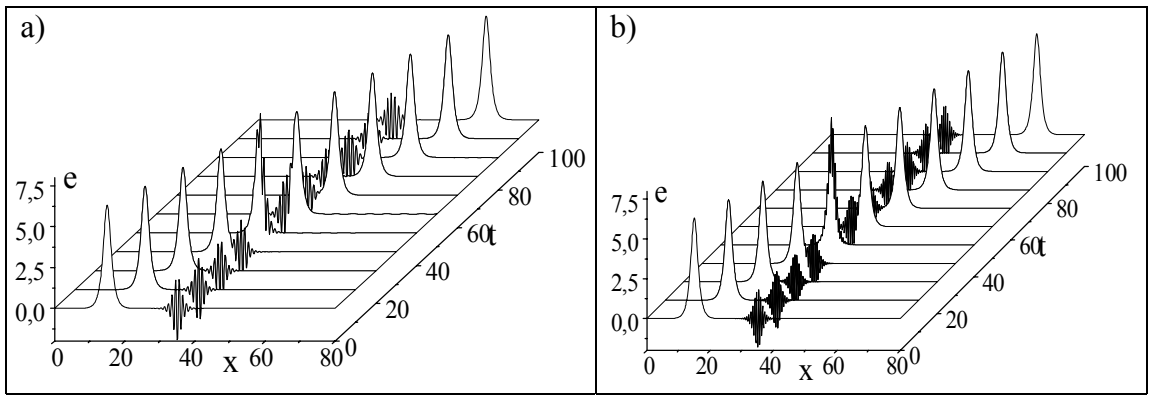

*Fig.* 12. Collision between a steady state pulse of the RMD model and a high-frequency pulse corresponding to mKdV breather. In (a) $\alpha_1=3$, and (b) - $\alpha_1=5$. The steady state pulse is characterized by $\alpha_1=3$. The parameter $\mu=0.1$.

In fig. 13 we show the evolution of a high-frequency pulse obtained by modulation of a steady state pulse of the RMD model by a harmonic wave.

As it appears, both an mKdV breather and a high-frequency pulse obtained by the modulation of a steady state pulse of the RMD model by a harmonic wave propagate steadily. Collision between these pulses and a steady state pulse do not lead to the decay of the breather-like pulses. The spectrum of the pulse around the frequency of the carrier wave is unchanged as the pulse propagates.

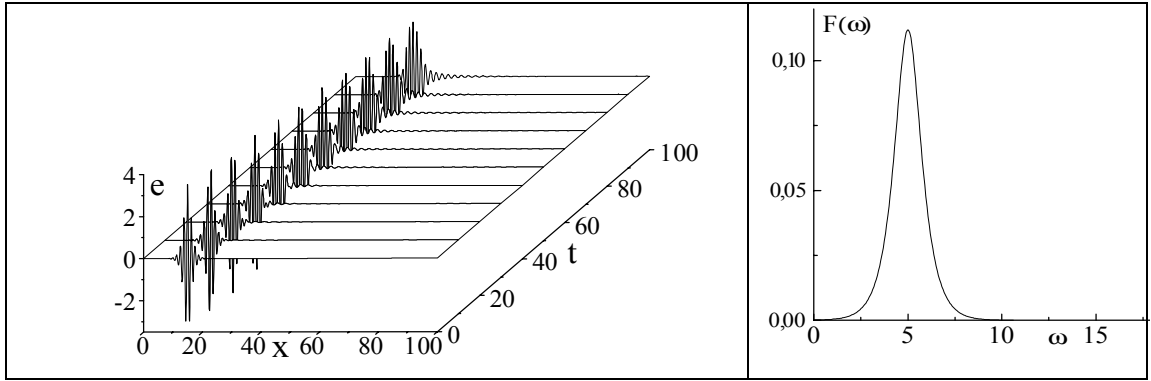

*Fig.* 13.

Left panel: evolution of the initial pulse
$$e(x, t=0) = (\alpha\sqrt{(\alpha-1)/\mu})\,\text{sech}[\sqrt{(\alpha-1)}(x-x_0)]\cos[5(x-x_0)] \quad \alpha = 2.$$
Right panel: Fourier spectrum of the pulse for the initial condition. It does not change with time. The parameter $\mu = 0.3$.

If we now reduce the frequency of the mKdV breather initial condition, it is quickly destroyed by dispersion and converted into quasi-harmonic periodic waves. Increasing the parameter corresponding to the breather frequency leads to stabilization of the pulse. Collisions of a steady state pulse with these high-frequency pulses also demonstrates the robustness of such quasi-breathers. In this way it should be concluded that high-frequency breather-like pulses in the RMD model are very close to the genuine breathers of a completely integrable system.

## 7. CONCLUSION

We have introduced and analyzed a model for the propagation of extremely short unipolar pulses of an electromagnetic field in a medium represented by anharmonic oscillators with a cubic nonlinearity. The model takes into account the dispersion properties of both the linear and nonlinear responses of the medium. It is the simplest generalization of the well-known Lorenz model used to describe linear optical properties in condensed matter. The cubic nonlinearity is the first type of anharmonic correction to the Lorenz model and it results in the Duffing oscillator. Here we consider the total Maxwell-Duffing model in detail. The Lagrangian picture of the RMD model was considered and three integrals of motion were found. Two families of exact analytical solutions, with positive and negative polarities, have been found as moving solitary pulses. The first kind of steady state ESP is an electromagnetic spike propagating in a nonlinear medium. It was discussed early in [1, 2, 6]. A new kind of steady state ESP is an electromagnetic spike propagating on a non-zero electric background. These can be both bright and dark ESP. Contrary to the ESP on a zero background, here pulses of different polarities have different amplitudes.

We found that the RMD equations can be represented in bilinear Hirota form. In the case of a zero background the one-soliton solution of the bilinear equations was obtained. It coincides with the expression of a steady state ESP. There are many examples when the bilinear form of nonlinear evolution equations leads to the existence of two-soliton solutions without having complete integrability and the existence of *N*-soliton solutions. However in this particular case we were not able to obtain a two-soliton solution of the RMD equations.

We investigated numerically the propagation of an ESP in the framework of the RMD model. This investigation demonstrates the stability of steady state solutions of RMD equations both on a zero background and a non-zero one.

**Acknowledgment**

Authors (A.I.M. and E.V.K.) are grateful to the *Laboratoire de Mathematiques, INSA de Rouen* for hospitality and support. The visit of Elena V. Kazantseva to Rouen was made possible due to support from F. Lederer and INTAS grant 96-339.